\documentclass[aps,prl,amssymb,amsmath,twocolumn,floatfix]{revtex4}
\usepackage{graphicx}
\usepackage{amsmath}

\begin{document}
\title{Influence of anisotropy on the study of critical behavior of spin models by machine learning methods}
\author{Diana Sukhoverkhova$^{1,2}$}
\author{Lev Shchur$^{1, 2}$}
\affiliation{$^1$ Landau Institute for Theoretical Physics, 142432 Chernogolovka, Russia}
\affiliation{$^2$ HSE University, 101000 Moscow, Russia}

\begin{abstract}
In this paper, we applied a deep neural network to study the issue of knowledge transferability between statistical mechanics models. The following computer experiment was conducted. A convolutional neural network was trained to solve the problem of binary classification of snapshots of the Ising model's spin configuration on a two-dimensional lattice. During testing, snapshots of the Ising model spins on a lattice with diagonal ferromagnetic and antiferromagnetic connections were fed to the input of the neural network. Estimates of the probability of samples belonging to the paramagnetic phase were obtained from the outputs of the tested network. The analysis of these probabilities allowed us to estimate the critical temperature and the critical correlation length exponent. It turned out that at weak anisotropy the neural network satisfactorily predicts the transition point and the value of the correlation length exponent. Strong anisotropy leads to a noticeable deviation of the predicted values from the precisely known ones. Qualitatively, strong anisotropy is associated with the presence of oscillations of the correlation function above the Stephenson disorder temperature and further approach to the point of the fully frustrated case.
\end{abstract}

\maketitle

\textbf{1. Introduction.} Machine learning has begun to form in the last decade an additional approach in the physical sciences in addition to the traditional approaches of experiment, theory, and computer modeling~\cite{Carleo-2019}. It seems appropriate to investigate this new approach from the point of view of applicability to certain problems and phenomena. It is also necessary to develop a methodology for estimating the quantities of interest with the necessary accuracy. In one of the first works on the application of machine learning to the study of phase transitions of the second kind, a method of supervised learning was proposed~\cite{Carrasquilla-2017} and satisfactory estimates of the critical temperature of model systems were obtained. This approach was complemented by a systematic method for estimating the critical exponent of correlation length~\cite{Chertenkov-2023}. The approach is based on binary classification using a deep neural network.

We raise the question of universality of application of the trained network. Training of a neural network is a rather costly process both in terms of forming a set of data for training and in terms of computer time and computer power. Neural networks have an advantage when using an already trained network, which is used, for example, for high-speed filtering of data from the Large Hadron Collider~\cite{Derkach-2018}. It is reasonable to study the versatility of using a neural network to extract information about the phase transition of the second kind. One of the first steps in this direction is to investigate the applicability of a neural network trained on the standard statistical mechanics model, the two-dimensional Ising model, for investigations of other models. For simplicity of formulation, testing was carried out on a model from the same universality class, but with diagonal anisotropy, which is known to have a nontrivial effect on the properties under study~\cite{Dohm-2021,Dohm-2023}.

A deep neural network (DNN)~\cite{Kumari-2017} is a multilayer network whose first layer is a matrix  $L{\times} L$. The layers can be fully connected or convolutional, each connection between layers has a weight~\cite{Carleo-2019}, which changes during training. A special feature of DNNs is the error back propagation mechanism, which allows the network to be trained efficiently~\cite{Rumelhart-1986} by changing the weights. In this paper, we use a convolutional neural network (CNN). The reader can find the details in~\cite{Chertenkov-2023}.

During {\em training}, the distribution of spins on the lattice $L{\times} L$ is fed to the input of the neural network and the belonging of each snapshot to the ferromagnetic or paramagnetic phase with respect to the known phase transition temperature is indicated. During {\em testing}, the trained network receives as input a snapshot of the lattice spins $L{\times} L$ at a known temperature $T$. At the output of one neuron, the network outputs $p_i{\in} [0,1]$ as an estimate of the paramagnetic phase membership of the sample under test. The other neuron outputs the belonging of the ferromagnetic phase $f_i{\equiv}{1{-}p_i}$. Therefore, in the following we will analyze only the paramagnetic phase belonging. Repeating the testing $N$ times at the given temperature of samples $T$ and lattice size $L$, we obtain an estimate of the probability $P(T;L)$ that the $N$ samples obtained in the simulation at temperature $T$ belong to the paramagnetic phase. The calculation of this function is performed by the formula~(\ref{PTL}). From the set of functions $P(T;L)$ the phase transition temperature will be determined using finite-dimensional analysis as proposed in~\cite{Carrasquilla-2017}. In addition, we will investigate the second moments of the probability distribution of belonging to the paramagnetic phase $D(T;L)$, formula~(\ref{DTL}). The study of the width and half-width of the Gaussian function approximating the distribution of $D(T;L)$, allows us to estimate the correlation length critical exponent~\cite{Chertenkov-2023}.

\begin{figure}
\center{\includegraphics[width=0.8\linewidth]{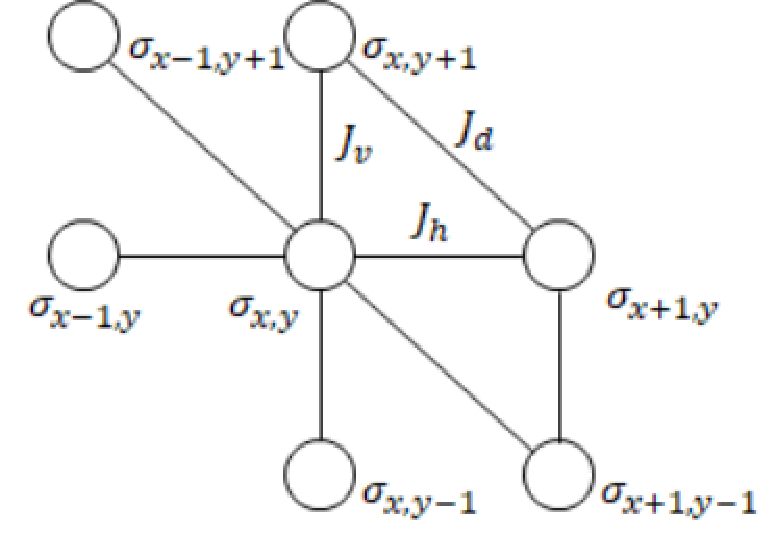}
\caption{Fig 1. Part of a lattice with a node $\sigma_{x,y}$ and its surroundings in the generalized Ising model.}
\label{fig:model}}
\end{figure}

\textbf{2. Models under study.} The Hamiltonian of the generalized Ising model has the form (Fig.~\ref{fig:model}):

\begin{equation}\label{eq:model}
    \mathcal{H}{=}{-}\sum_{x,y=1}^L\sigma_{x,y}\left[J_h\sigma_{x+1,y}{+}J_v\sigma_{x,y+1}{+}J_d\sigma_{x+1,y+1}\right].
\end{equation} 
The phase transition temperature of this model is known from the analytical solution~\cite{Houtappel-1950}
\begin{equation}\label{eq:tc}
\begin{split}
    \sinh{\frac{2J_v}{k_BT_c}}\; \sinh{\frac{2J_h}{k_BT_c}}+\sinh{\frac{2J_h}{k_BT_c}}\; \sinh{\frac{2J_d}{k_BT_c}}+\\+\sinh{\frac{2J_d}{k_BT_c}}\; \sinh{\frac{2J_v}{k_BT_c}} = 1 \\
    J_v+J_h>0,~J_h+J_d>0,~J_d+J_v>0
\end{split}
\end{equation}

We will train on the isotropic model at the value of parameters $J_d{=}0,~J_h{=}J_v{=}1$. Testing will be carried out for the case of diagonal anisotropy with a set of parameters $J_h{=}J_v{=}J,~J_d{\neq} 0$. The constant $J_d$ can be either positive or negative, bounded by the inequality $J_d{+}J{>}0$.

Given a given set of temperature values, the Metropolis algorithm generates a sample of $N$ uncorrelated configurations of spins. The thermalization time is $20{\times} L^{2.15}$ as estimated by~\cite{Chertenkov-2021}, and configurations are stored (snapshot!) once in $2{\times} L^{2.15}$ Monte Carlo steps, measured in the number $L^2$ of local flips of the lattice spins $L{\times}L$.

\textbf{3. Neural network training.} We used a convolutional neural network (CNN) architecture with one convolutional, two fully connected layers and ReLU activation between them~\cite{Chertenkov-2023}. The network was trained on uncorrelated images from simulations of the isotropic Ising model, $J_d{=}0,~J_h{=}J_v{=}1$, with linear dimensions of the lattices $L{=}20,30,60,80,100,120$. At each size $L$ we selected $100$ temperature values in increments $0.006$ from the temperature range: 
\begin{equation}
T{\in} [T_c{-} 0.3,T_c  {+} 0.3],
\label{eq-T}
\end{equation}
where the critical temperature was determined by the relation~({\ref{eq:tc}). The number of snapshots during training at each temperature $T$ and each size $L$ was chosen $M{=}2048$.

The network was trained for each linear lattice size $L$ for one epoch to avoid overtraining of the network. When the number of epochs is exceeds $20$, overtraining leads to an almost unambiguous classification into ferromagnetic and paramagnetic phases with a sufficiently accurate estimate of the transition temperature, but the information on the properties of the correlation length is completely lost~\cite{Chertenkov-diss}.

\textbf{4. Testing.} Testing was performed on $N{=}512$ snapshots obtained from the simulation for given coupling constants, at each value of $T$ and $L$, with the same $L$ values as in training. The temperature values during testing were also selected from the range~(\ref{eq-T}) relative to the critical temperature calculated by formula~(\ref{eq:tc}). This choice naturally includes the temperature range of interest in the vicinity of the phase transition of the model under test. 

As noted in the introduction, each of the $N$ samples tested is given an estimate $p_i$ of paramagnetic phase membership. For each temperature value and for each linear size of the lattice, we calculate an estimate of the probability that the sample belongs to the paramagnetic phase
\begin{equation}
    P(T;L){=}\frac{1}{N}\sum_{i=1}^N p_i(T;L)
\label{PTL}
\end{equation}
and the corresponding standard deviations
\begin{equation}
    D(T;L){=}\sqrt{\dfrac{1}{N}\sum\limits_{i=1}^N(p_i(T;L))^2{-}\left(\dfrac{1}{N}\sum\limits_{i=1}^N p_i(T;L)\right)^2}
    \label{DTL}.
\end{equation}

\begin{figure}
\center
\includegraphics[width=0.8\linewidth]{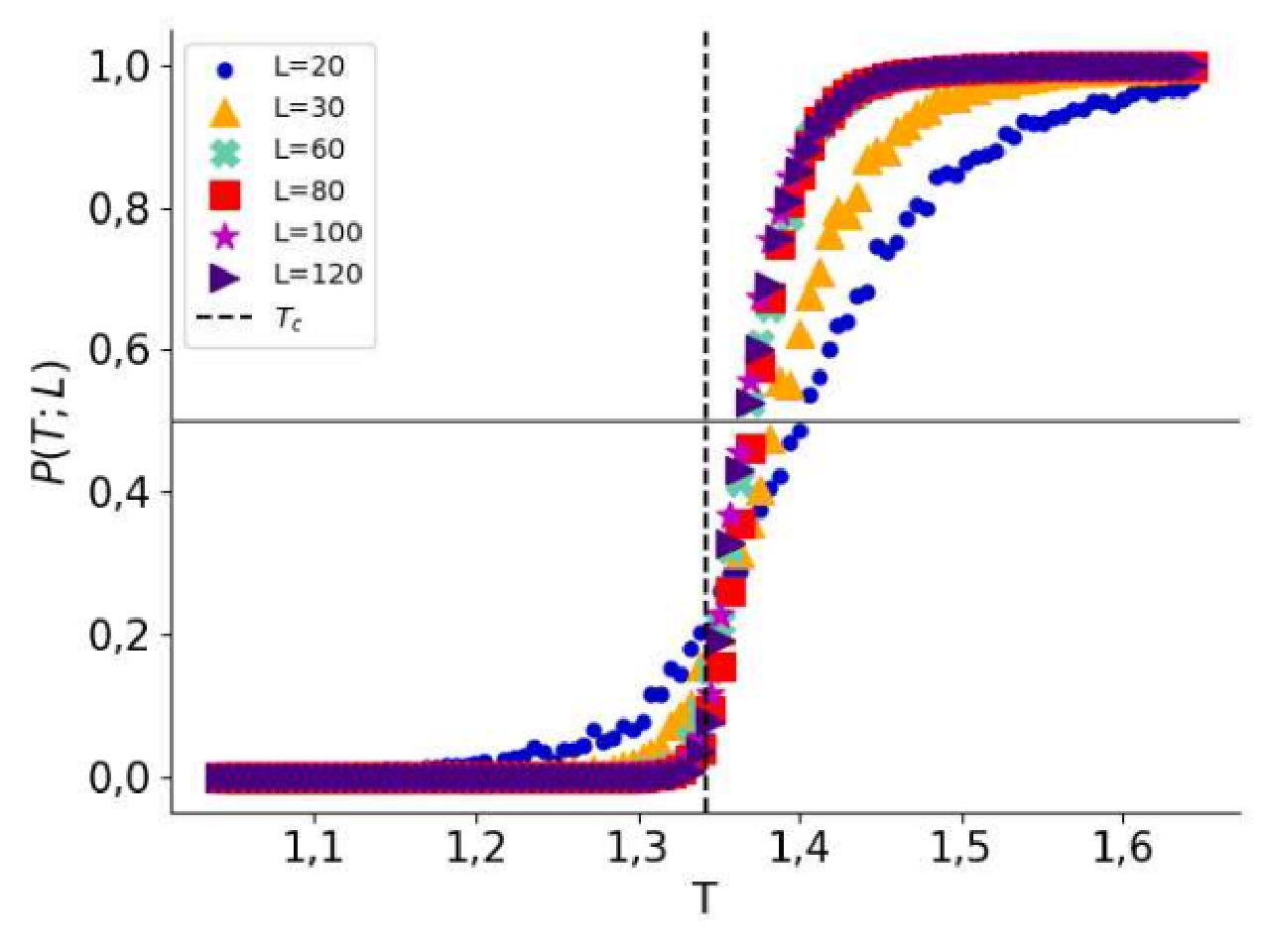}
\caption{Fig. 2. (Color online) Estimation of paramagnetic phase membership $P(T;L)$ (\ref{PTL}) of the Ising model with diagonal anisotropy at the value of the anisotropy parameter $J_d/J{=}{-}0.5$.}
\label{fig:PTL-example}
\end{figure}

\begin{figure}
\center
\includegraphics[width=0.8\linewidth]{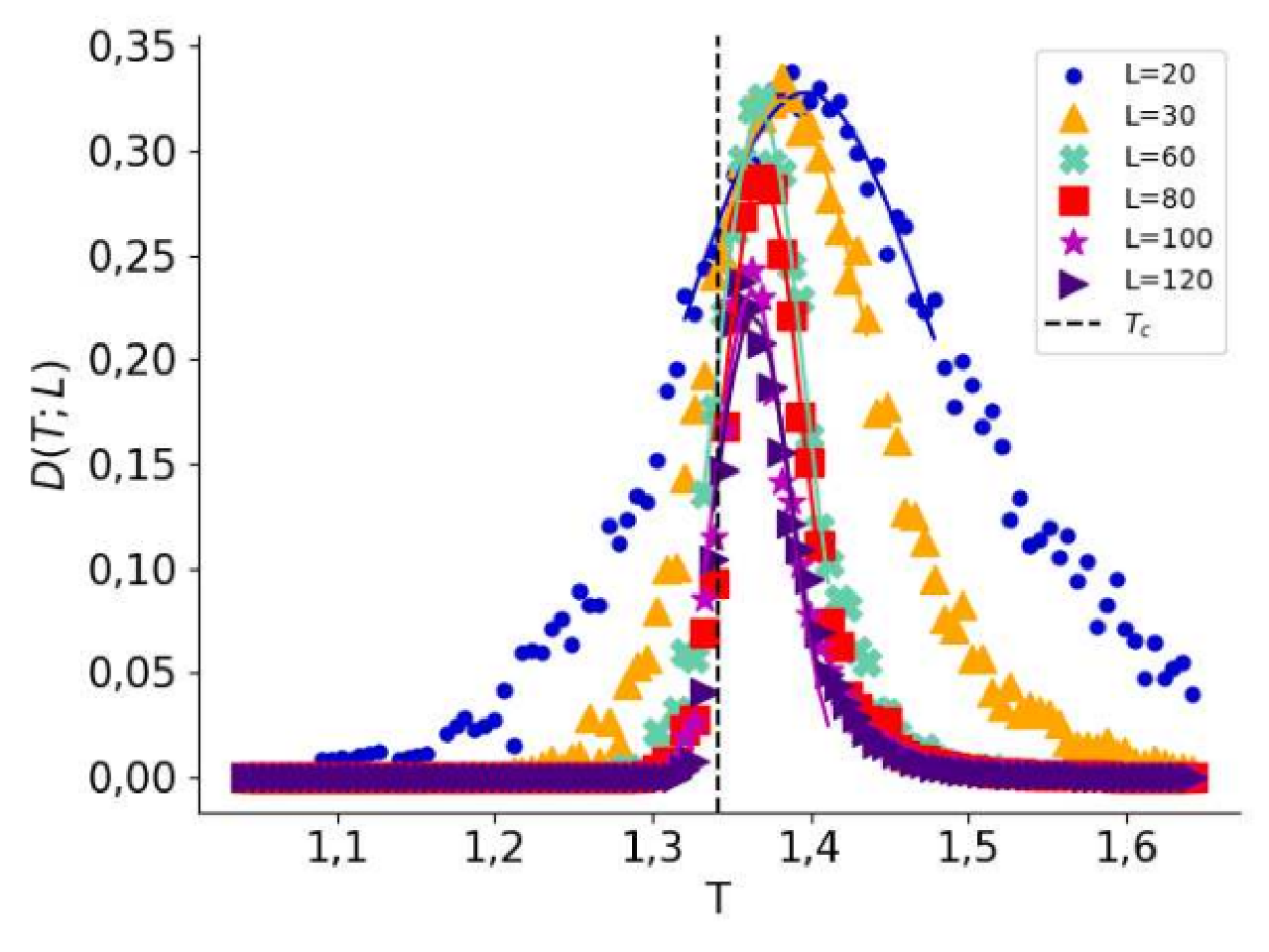}
\caption{Fig. 3. (Color online) Standard deviation function of paramagnetic phase membership $D(T;L)$ of the Ising model with diagonal anisotropy, $J_d/J{=}{-}0.5$}
\label{fig:DTL-example}
\end{figure}

\textbf{5. Estimation of critical temperature.} In~\cite{Carrasquilla-2017} was proposed a method for estimating the critical temperature based on analyzing the $P(T;L)$ on the basis of the hypothesis of correspondence of pseudocritical temperature $T^*_c$ to the value of the probability estimate $P(T^*_c;L){=}0.5$. In Fig.~\ref{fig:PTL-example} the result of calculation of the set of functions $P(T;L)$ with antiferromagnetic interaction along the diagonal $J_d{<}0$ at the value of the anisotropy parameter $J_d/J{=}{-}0.5$ is given. The corresponding $D(T;L)$ functions are shown in Fig.~\ref{fig:DTL-example}.

We used the following averaged estimates of the probabilities $P(T;L)$ to obtain an estimate of the critical temperature:

\begin{enumerate}
    \item The temperature at which the network prediction is closest to $0.5$:
    \begin{equation}\label{eq:tc_star}
        T^{*}_c(L) {=} \min_{T}|P(T;L){-}0.5|
    \end{equation}
    \item The temperature corresponding to the point of intersection of the line connecting the two closest to $0.5$ network predictions with $y{=}0.5$:
    \begin{equation*}
        \begin{split}
        T_{min} = \min_{T}|P(T;L)-0.5|,~P(T;L)<0.5 \\
        T_{max} = \min_{T}|P(T;L)-0.5|,~P(T;L)>0.5 \\
        P_{min}=P(T_{min};L);~P_{max}=P(T_{max};L)         
        \end{split}
    \end{equation*}    
    \begin{equation}
        T^{\circ}_c(L) {=} \frac{T_{min}(P_{max}-0.5)+T_{max}(0.5-P_{min})}{P_{max}-P_{min}}  
        \label{TC0}
    \end{equation}
    \item  Temperature based on the function $D(T;L)$: the distribution $D(T;L)$ at each $L$ is approximated by a Gaussian function with mean $\mu$ and standard deviation $\sigma$. We consider $\mu(L)$ as an estimate of the critical temperature.
\end{enumerate}

For each set, we calculated the asymptotic value of the critical temperature estimate using the well-known expression for the finite-dimensional shift of the pseudocritical temperature~\cite{Fisher-1967,Ferdinand-1969}. For example, for the variant 1 estimate, this can be written as
\begin{equation}
    T^*_c{=}T^*_c(L ){+}a/L\label{eq-FF}.
\end{equation}
The deviation of the obtained linear approximation of the $T^*_c(L )$ points with respect to the exact critical temperature is presented in Fig.~\ref{fig:jd-tcs} by black squares together with the statistical error. The critical temperature estimates for the remaining two cases are calculated in a similar manner. Note that all three methods of critical temperature estimation give statistically equivalent results. The deviation of the critical temperature estimate from a precisely known value is discussed in Section 7 of the paper.

\begin{figure}
\center
\includegraphics[width=\linewidth]{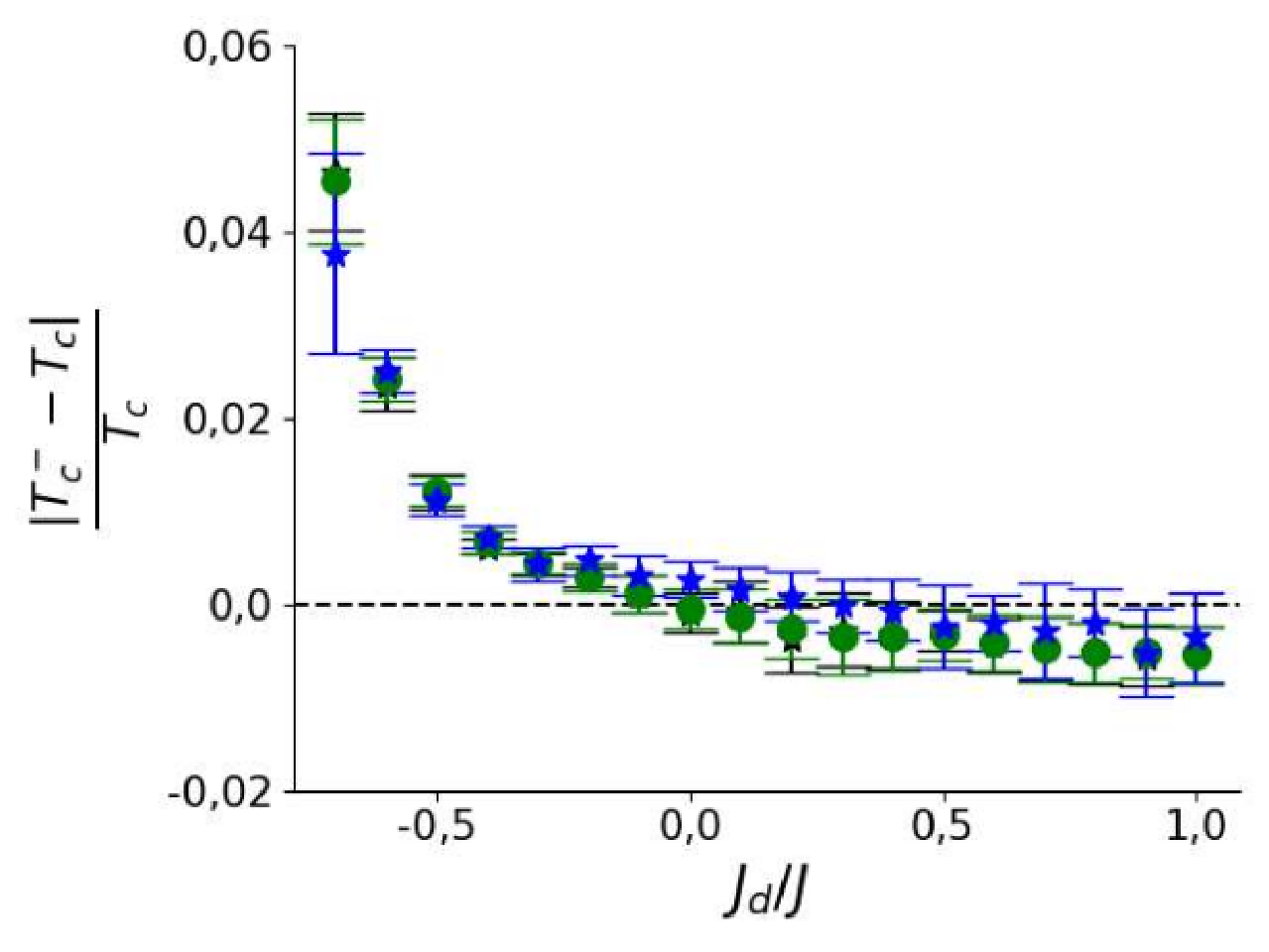}
\caption{Fig. 4. (Color online) Relative error of critical temperature estimation $T^-_c$ in the limit of infinite lattice size as a function of the anisotropy parameter $J_d{/}J$. The symbols correspond to the three ways of estimating the pseudocritical temperature indicated in the text: black squares – $T^*_c$; green circles – $T^{\circ}_c$; blue stars – $\mu$. $J_d/J$.}
\label{fig:jd-tcs}
\end{figure}

\textbf{6. Estimation of the correlation length critical exponent.} The correlation length $\xi$ in the Ising model~(\ref{eq:model}) in the phase transition region grows according to the $\xi {\propto} \tau^{-1/\nu}$ power law with decreasing reduced temperature $\tau{=}(T {-} T_c)/T_c$. The critical exponent $\nu$ is equal to one~\cite{Onsager-1941}.

With $L$ fixed, we approximate each function $D(T;L)$ by a Gaussian, discarding statistically insignificant values at the edges of the temperature range, and analyze the resulting $\mu(L)$ values as described in (3) of the previous section to estimate the critical temperature as well as the width $\sigma(L)$. Here we use the hypothesis~\cite{Chertenkov-2023}, that the width $\sigma(L)$ scales in the same way as some thermodynamic functions~\cite{Fisher-1967,Ferdinand-1969}, with the correlation length exponent $\nu$:
\begin{equation}
    \sigma(L){\propto}\frac{b}{L^{1/\nu}} \label{eq-FF2}.
\end{equation}

This hypothesis was confirmed in the paper~\cite{Chertenkov-2023} for models in two universality classes, in which reliable estimates of the correlation length exponent were obtained for the two-dimensional Ising model on a square lattice and the Baxter-Wu model formulated on a triangular lattice.

The results of the estimation of the critical correlation length exponent are depicted in Fig.~\ref{fig:jd-nu}, plotted against the data in Table~\ref{table:jd-nu}.

\begin{figure}
\center{\includegraphics[width=\linewidth]{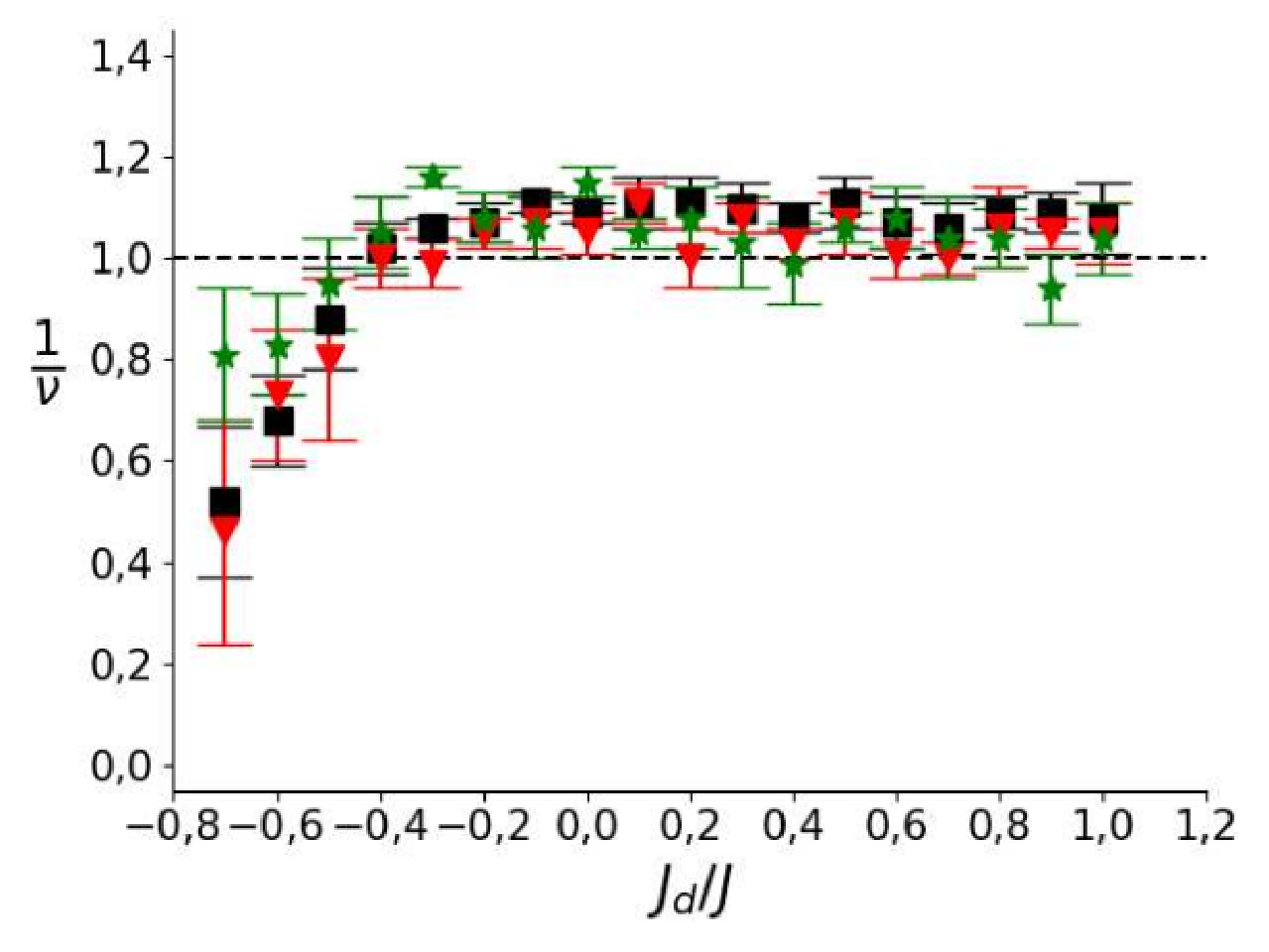}
\caption{Fig. 5. (Color online) Estimation of the inverse of the critical correlation length exponent: black squares are full-width estimates, red triangles by the right side, green stars by the left side.}
\label{fig:jd-nu}}
\end{figure}

\begin{table}[h!]
\center
\begin{tabular}{|c|c|c|c|c}				\cline{1-4}
    $J_d/J$  & $\frac{1}{\nu}$    & $(\frac{1}{\nu})_r$ & $(\frac{1}{\nu})_l$ &  \\ \cline{1-4}
    \textbf{-0.7} & 0.52(15) & 0.46(22) & 0.81(13)   &  \\ \cline{1-4}
    \textbf{-0.6} & 0.68(9)  & 0.73(13)  & 0.83(10)   &  \\ \cline{1-4}
    \textbf{-0.5} & 0.88(10)  & 0.80(16)  & 0.95(9)  &  \\ \cline{1-4}
    \textbf{-0.4} & 1.02(5)  & 1.00(6)  & 1.05(7)   &  \\ \cline{1-4}
    \textbf{-0.3} & 1.06(2)  & 0.99(5)  & 1.16(2)  &  \\ \cline{1-4}
    \textbf{-0.2} & 1.07(4)  & 1.05(3)  & 1.08(5)   &  \\ \cline{1-4}
    \textbf{-0.1} & 1.11(2)  & 1.07(5)  & 1.06(6)  &  \\ \cline{1-4}
    \textbf{0.0} & 1.09(2)  & 1.05(4)  & 1.15(3)  &  \\ \cline{1-4}
    \textbf{0.1} & 1.11(5)  & 1.11(4)  & 1.05(3)  &  \\ \cline{1-4}
    \textbf{0.2} & 1.11(5)  & 1.00(6)  & 1.08(6)   &  \\ \cline{1-4}
    \textbf{0.3} & 1.10(5)  & 1.08(3)  & 1.03(9)  &  \\ \cline{1-4}
    \textbf{0.4} & 1.08(3)  & 1.03(3)  & 0.99(8)   &  \\ \cline{1-4}
    \textbf{0.5} & 1.11(5)  & 1.07(6)  & 1.06(3)  &  \\ \cline{1-4}
    \textbf{0.6} & 1.07(5)  & 1.01(5)  & 1.08(6)   &  \\ \cline{1-4}
    \textbf{0.7} & 1.06(5) & 1.00(3) & 1.04(8)   &  \\ \cline{1-4}
    \textbf{0.8} & 1.09(3) & 1.06(8) & 1.04(6)   &  \\ \cline{1-4}
    \textbf{0.9} & 1.09(4) & 1.05(3) & 0.94(7)   &  \\ \cline{1-4}
    \textbf{1.0} & 1.08(7) & 1.05(6) & 1.04(7)   &  \\ \cline{1-4}
\end{tabular}
\caption{Table 1. Estimates of the inverse of the critical correlation length exponent $\nu$, obtained from analyzing the width $\sigma(L)$ of the function $D(T;L)$ – second column, the right half-width $\sigma_r(L)$ of this function – third column and the left half-width $\sigma_l(L)$ of this function – fourth column.}
\label{table:jd-nu}
\end{table}

\textbf{7. Discussion.} We used a convolutional neural network to solve the binary classification problem of {\em isotropic} Ising model. By analyzing the predictions of the neural network, we determined the phase transition temperature $T_c$ and correlation length exponent $\nu$ with good agreement~\cite{Carrasquilla-2017,Chertenkov-2023} with the exact analytical result~\cite{Onsager-1941}.

We used the trained network to test the Ising model with {\em diagonal anisotropy} with ferromagnetic and antiferromagnetic bonds. In the region of moderate values of the anisotropy parameter, the network trained on the isotropic model predicts the phase transition temperature and the value of the correlation length exponent within statistical error, which can be seen from Table~\ref{table:jd-nu} and Figs.~\ref{fig:jd-tcs} and~\ref{fig:jd-nu} for values of the anisotropy parameter $J_d/J$ from ferromagnetic anisotropy at $J_d/J{=}1$ to antiferromagnetic anisotropy at $J_d/J{=}{-}0.4$.

At the values of the anisotropy parameter $J_d/J{<}{-}0.4$, a systematic deviation of both the transition temperature and the correlation length exponent from the exact values is observed. This deviation can be explained by the peculiarity of the phase diagram of the model studied by Stephenson~\cite{Stephenson-1970} and the presence in the temperature range under test of snapshots of the spin distribution falling above the temperature, which in the paper~\cite{Stephenson-1970} was called disorder temperature $T_D$, above which there is a region with oscillating pair correlations of spins. The oscillations are due to the competing diagonal antiferromagnetic interaction. Since such oscillations were completely absent when training the neural network, the test results lead to deviations in the estimates of the critical temperature and the correlation length exponent.

Note that diagonal anisotropy also affects the observed thermodynamic quantities in a nontrivial way. For example, it was predicted analytically~\cite{Chen-2004} and found numerically~\cite{Selke-2005} that the value of the Binder cumulant, constructed from the ratio of the square of the second magnetization moment to the fourth magnetization moment, varies depending on the diagonal anisotropy parameter. Previously, its value at the phase transition point was thought to be determined solely by the universality class of the model. The variation of the Binder cumulant at positive values of the anisotropy parameter is small and amounts to less than a percent, as shown in Fig.~4 from the paper~\cite{Selke-2009}. One can see a sharp decline at values of the anisotropy parameter $-0.4$, similar to our cross-training/testing data.

Thus, we have established qualitatively the limits of applicability of cross-training in testing by a neural network of a model with diagonal anisotropy and antiferromagnetic interaction. These limits are related to the peculiarities of the phase diagram of the anisotropic model~\cite{Stephenson-1970} and the proximity to the point of transition of the system to a fully frustrated state at $J_d/J{=}{-}1$, near which there exist long-lived topological excitations studied in the works of~\cite{Korshunov-2005,Smerald-2016}.

\textbf{Funding.} This work was supported by the Russian Scientific Foundation grant 22-11-00259.

The supercomputer complex of the National Research University Higher School of Economics was used for computer modeling and machine learning.

\textbf{Conflict of interest.} The authors of this work declare that they have no conflicts of interest.


\begin{thebibliography}{99}

\bibitem{Carleo-2019} G. Carleo, I. Cirac, K. Cranmer, L. Daudet, M. Schuld, N. Tishby, L. Vogt-Maranto, and L. Zdeborová, Rev. Mod. Phys. {\bf 91},  045002 (2019).

\bibitem{Carrasquilla-2017} J. Carrasquilla and R.G. Melko, Nat. Phys. {\bf 13}, 431 (2017).

\bibitem{Chertenkov-2023} V. Chertenkov, E. Burovski,  and L. Shchur, Phys. Rev. E {\bf 108}, L031102 (2023).

\bibitem{Derkach-2018} D. Derkach, et al., J. Phys.: Conf. Ser. {\bf 1085}, 042038 (2018).

\bibitem{Dohm-2021} V. Dohm and S. Wessel, Phys. Rev. Lett. {\bf 126}, 060601 (2021).

\bibitem{Dohm-2023} V. Dohm, Phys. Rev. E {\bf 108}, 044149 (2023).

\bibitem{Kumari-2017} R. Kumari and S.K. Srivastava, International Journal of Computer Applications {\bf 160}, 11 (2017).

\bibitem{Rumelhart-1986} D.E. Rumelhart, G.E. Hinton, and R.J. Williams, Nature {\bf 323.6088}, 533 (1986).

\bibitem{Houtappel-1950} R.M.F. Houtappel, Physica {\bf 16}, 425 (1950).

\bibitem{Chertenkov-2021} V. Chertenkov and L. Shchur, Journal of Physics: Conference Series {\bf 740}, 012003 (2021).

\bibitem{Chertenkov-diss} V. I. Chertenkov, PhD thesis, National Research University Higher School of Economics (2024).

\bibitem{Fisher-1967} M.E Fisher and A.E. Ferdinand, Phys. Rev. Lett. {\bf 19}, 169 (1967).

\bibitem{Ferdinand-1969} A.E. Ferdinand and M.E. Fisher, Phys. Rev. B {\bf 185}, 832 (1969).

\bibitem{Onsager-1941} L. Onsager, Phys. Rev. {\bf 65}, 117 (1941). 

\bibitem{Stephenson-1970} J. Stephenson, Phys. Rev. B {\bf 1}, 4405 (1970).

\bibitem{Chen-2004} X.S. Chen and V. Dohm, Phys. Rev. E {\bf 70}, 056136 (2005).

\bibitem{Selke-2005} W. Selke and L.N. Shchur, J. Phys. A: Math. Gen. {\bf 38}, L739 (2005).

\bibitem{Selke-2009} W. Selke and L.N. Shchur, Phys. Rev. E {\bf 80}, 042104 (2009).

\bibitem{Korshunov-2005} S.E. Korshunov, Phys. Rev. B {\bf 72}, 144417 (2005).

\bibitem{Smerald-2016} A. Smerald, S. Korshunov, and F. Mila, Phys. Rev. Lett. {\bf 116}, 197201 (2016).


\end{thebibliography}
\end{document}